\documentclass{amsart}
\usepackage{amssymb}
\usepackage{latexsym}

\newtheorem{theorem}{Theorem}
\newtheorem{lemma}{Lemma}[section]
\newtheorem{prop}[lemma]{Proposition}
\newtheorem{coro}[lemma]{Corollary}
\newtheorem{definition}[lemma]{Definition}

\newcommand{\R}{{\mathbb R}}
\newcommand{\C}{{\mathbb C}}
\newcommand{\N}{{\mathbb N}}

\newcommand{\Z}{{\mathbb Z}}
\newcommand{\E}{{\mathbb E}}

\newcommand{\Cplus}{{\mathbb C}_+}

\newcommand{\unull}{\mbox{u}_{0,z}}
\newcommand{\ueins}{ \mbox{u}_{1,z} }

\newcommand{\uinf}{ \mbox{u}_{\infty,z} }

\newcommand{\uunull}{ \mbox{\underline{u}}_{0,z} }
\newcommand{\uueins}{ \mbox{\underline{u}}_{1,z} }
\newcommand{\uuf}{ \mbox{\underline{u}}_{f,z} }

\newcommand{\Tjt}{\widetilde{T}_j}
\newcommand{\ettheta}{\widetilde{e}_\theta}
\newcommand{\Seins}{\R / 2 \pi \Z }
\newcommand{\Sdeltaj}{\mathcal{S}_{\delta,j}}
\newcommand{\Smitallem}{\mathcal{S}_{\delta,\omega}^{k,m}}
\newcommand{\Smitallemn}{\mathcal{S}_{\delta,\omega}^{k}}

\renewcommand{\Re}{{\rm Re}}
\newcommand{\tr}{{\mathrm{tr}}}

\newcounter{smalllist}

\begin{document}
\title[Lower transport bounds for 1d Schr\"odinger operators]{Lower
transport bounds for one-dimensional continuum Schr\"odinger
operators}
\author[D.~Damanik, D.~Lenz, G.~Stolz]
{David Damanik$\,^{1}$, Daniel Lenz$\,^{2}$, G\"unter Stolz$\,^{3}$,   }

\thanks{D.\ D.\ was supported in part by NSF grant DMS--0227289}

\thanks{G.\ S.\ was supported in part by NSF grant DMS--0245210}

\maketitle

\noindent $^1$ Mathematics 253--37, California Institute of Technology,
Pasadena, CA 91125, U.S.A., E-Mail: damanik@caltech.edu\\[0.2cm]
$^2$ Fakult\"at f\"ur Mathematik, TU Chemnitz, D-09107 Chemnitz, Germany, E-Mail:
dlenz@mathematik.tu-chemnitz.de\\[0.2cm]
$^3$ Department of Mathematics, University of Alabama, Birmingham, AL 35294, U.S.A.,
E-Mail: stolz@math.uab.edu

\bigskip

\begin{center}
\textit{Dedicated to Joachim Weidmann on the occasion of his 65th
birthday.}
\end{center}

\medskip

\begin{abstract}
We prove quantum dynamical lower bounds for one-dimensional continuum Schr\"odinger
operators that possess critical energies for which there is slow growth of transfer
matrix norms and a large class of compactly supported initial states. This general result
is applied to a number of models, including the Bernoulli-Anderson model with a constant
single-site potential.
\end{abstract}

\section{Introduction}

We study one-dimensional Schr\"odinger operators associated to
\begin{equation} \label{eq1.1}
\varphi \: \mapsto \: -\varphi''+V\varphi,
\end{equation}
where $V: \R \to \R$ satisfies
\begin{equation} \label{eq1.2}
\|V\|_{1,\mbox{\footnotesize unif}} := \sup_{x\in\R}
\int_{x}^{x+1} |V(t)|\,dt < \infty.
\end{equation}

We will have results for the associated whole-line operator, denoted by $H$, as well as
for the associated selfadjoint operator $H_D$ on $[0,\infty)$ with Dirichlet boundary
condition at $0$ (which could be easily adjusted to other boundary conditions).

We are interested in situations where non-trivial quantum transport for systems governed
by the above Hamiltonians can be established. To this end we will consider the time
averaged $p$-th moments of the position operator $(X\varphi)(x) = x\varphi(x)$ with given
initial state $f$:
\begin{equation} \label{eq1.3}
M_f(T,p) := \frac{2}{T} \int_0^{\infty}
\exp\left(-\frac{2t}{T}\right) \left\| |X|^{p/2} \exp(-itH)f
\right\|^2 \,dt.
\end{equation}
In the same way we define $M_{f,D}(T,p)$ if $H$ is replaced by $H_D$. The presence of
transport will be proven through lower bounds for the lower growth exponents (lower {\em
diffusion exponents})
\begin{equation} \label{eq1.4}
\beta_f^-(p) := \liminf_{T\to\infty} \frac{\log M_f(T,p)}{\log T},
\end{equation}
and similarly $\beta_{f,D}^-(p)$.

Using Abelian means in (\ref{eq1.3}) is convenient for our proofs.
This is done by most authors and may in most applications be
replaced by Cesaro means
\begin{equation} \label{eq1.5}
\frac{1}{T} \int_0^T \left\| |X|^{p/2} \exp(-itH)f \right\|^2
\,dt,
\end{equation}
without changing the value of the diffusion exponents. This is
easy to see if an a-priori upper bound $\| |X|^{p/2} \exp(-itH)f
\|^2 \le C|t|^N$ is available. The latter arises for example in
the form of ballistic upper bounds on quantum transport (i.e.\
$N=p$), which hold in great generality, see e.g.\
\cite{Radin/Simon} for $p=2$ and $p=4$.

For discrete one-dimensional Schr\"odinger operators on $\ell^2(\N)$ and $\ell^2(\Z)$,
respectively, Damanik and Tcheremchantsev \cite{DT} have developed a general method which
allows one to derive lower bounds on diffusion exponents from upper bounds on the growth
of norms of transfer matrices. In Section~\ref{s2} we will present an extension of their
method to continuum operators. Due to our intended applications we will focus on results
which arise from transfer matrix bounds in the vicinity of a single ``critical'' energy.

The most interesting issue which arises in this extension is the
question for the proper choice of the initial state $f$. The paper
\cite{DT} only considers the case $f= \delta_1$, a discrete unit
mass. While this is somewhat natural in discrete space, there is
no corresponding choice in the continuum (at least if one wants to
stay in Hilbert space). Our results will allow for any compactly
supported initial state as long as it ``feels'' the critical
energy in the sense that it is not orthogonal to the
eigensolutions of the Schr\"odinger equation at this energy.

Our main motivation for extending the methods of \cite{DT} to the continuum comes from
applications to random Schr\"odinger operators, specifically continuum Bernoulli-Anderson
models where the random coupling constants take only two possible values. These operators
are known to almost surely exhibit {\em spectral localization}, that is they have pure
point spectrum with exponentially decaying eigenfunctions \cite{DSS}. On the other hand,
it was also observed in \cite{DSS} that these operators may have a discrete set of
critical energies at which the Lyapunov exponent vanishes. Dynamical localization (in the
sense of time-boundedness of all moments of the position operator) was obtained in
\cite{DSS} only after projecting onto energy intervals which have positive distance from
the critical energies.

In Section~4 we will show that the existence of critical energies in continuum
Bernoulli-Anderson models indeed gives rise to quantum transport in the sense that almost
surely
\begin{equation} \label{eq1.7}
\beta_f^-(p) \ge p - \frac{1}{2}
\end{equation}
for all $p>0$ and suitable $f$.

As a prototype of a continuum Bernoulli-Anderson model consider
\begin{equation} \label{eq1.6}
-\frac{d^2}{dx^2} + \sum_{n\in\Z} \omega_n \chi_{[n,n+1]},
\end{equation}
where the i.i.d.\ random variables $\omega_n$ only take the values $0$ or $1$. This
operator has not just one but infinitely many critical energies at which the Lyapunov
exponent vanishes. We find that \eqref{eq1.7} holds almost surely for all
square-integrable $f$ with support in $[0,1]$.

Thus \eqref{eq1.6} provides a model for the co-existence of spectral localization and
dynamical delocalization in the form of super-diffusive transport. The latter is best
characterized by the mean-square deviation, i.e.\ $p=2$ in \eqref{eq1.3}: $\beta(2)=2$
would be ballistic transport, $\beta(2)=1$ diffusive, while we get $\beta(2) \ge 3/2$.
Our work provides a continuum analogue of results established previously for the discrete
{\em dimer model} (see \cite{BG} for spectral localization and \cite{JSBS} for the lower
transport bound \eqref{eq1.7}).

We point out that the co-existence phenomenon arises for continuum Anderson models
already in the prototypical case \eqref{eq1.6}, while in the discrete case the standard
Anderson model (with independent sites) has no critical energies and is spectrally and
dynamically localized for any non-trivial distribution of the single site couplings
\cite{CKM,dBG}.

Obtaining the probabilistic transfer matrix bounds which are necessary to deduce
\eqref{eq1.7} for Bernoulli-Anderson models is quite subtle (while $\beta_f^-(p) \ge p-1$
follows from a much simpler deterministic bound). To get \eqref{eq1.7}, which is
physically expected to be the exact diffusion exponent (at least for the dimer and $p=2$,
see \cite{DWP}), we employ a law of large numbers type result from \cite{JSBS}.

In Section~5 we add another application of our general results in Section~2. Here we
consider self-similar potentials which are generated by means of a substitution rule. We
discuss potentials generated by the Thue-Morse substitution or the period doubling
substitution in detail and prove the existence of critical energies at which the norms of
transfer matrices remain uniformly (resp., linearly) bounded. This then yields the lower
bound $p-1$ (resp., $(p-5)/2$) for diffusion exponents.

We conclude this introduction by comparing our approach to dynamical lower bounds for
continuum Schr\"odinger operators with previous ones.

The first general method to prove dynamical lower bounds for Schr\"odinger operators with
singular spectral measures goes back to Guarneri \cite{G} and was further developed by
Combes \cite{C} and Last \cite{L}. Dynamical lower bounds are found in terms of
continuity properties of spectral measures with respect to Hausdorff measures. While this
correspondence holds in arbitrary dimension, this approach is particularly useful in one
dimension since the required input can be established using the Jitomirskaya-Last
extension \cite{JL1,JL2} of Gilbert-Pearson theory \cite{GP}. Nevertheless, proofs of
Hausdorff-absolute continuity of spectral measures are often quite involved or even
impossible. For example, within the class of self-similar potentials, only (discrete)
potentials of Fibonacci type could be handled, models associated with Thue-Morse or
period doubling symmetry are as yet outside the scope of this approach. Moreover, the
required spectral continuity may not hold at all for interesting models. For example, the
Bernoulli-Anderson model has pure point spectrum and hence no useful spectral continuity
properties.

Another method was recently developed by Germinet, Kiselev, and Tcheremchantsev
\cite{GKT}. While their approach is similar in spirit to ours, namely that upper bounds
on transfer matrix norms imply lower bounds for diffusion exponents, our results give
better bounds for the applications we have in mind. Their method is particularly suitable
for models that admit power-law upper bounds on transfer matrix norms for large sets of
energies, and hence their applications establish good dynamical bounds for models with
this feature, such as random decaying potentials and sparse potentials. For models with
small (e.g., finite) sets of such energies, our method gives better dynamical bounds. For
example, their method combined with our Theorem~\ref{dreiviertelschranke} below gives the
bound $\beta_f^-(p) \ge \frac{1}{2} (p - 1)$ in the case of the Bernoulli-Anderson model
with a critical energy, whereas we obtain the stronger bound \eqref{eq1.7}, which is
conjectured to be optimal. Moreover, their proof of the general dynamical bound is more
involved than ours. Thus, while the results of \cite{GKT} and this paper are somewhat
related, the scopes in terms of applications are almost disjoint.

\medskip

\noindent {\bf Acknowledgement.} D.\ L.\ gratefully acknowledges visits to Caltech and
UAB during which part of this work was done.

\section{The main result}   \label{s2}

In this section, we develop the continuum analog of the approach
to quantum dynamical lower bounds from \cite{DT}. As discussed in
the introduction, working in the continuum requires to come up
with a better understanding of which initial states will generate
transport. The key technical input which settles this issue is
contained in Lemma~\ref{crucial}. As a bonus, this observation
also suggests how the dynamical lower bound in the discrete case
can be extended to arbitrary (finitely supported) initial states.
We will discuss this extension in Section~\ref{Remarks}.

A central role will be played by solutions of
\begin{equation}\label{homogen} -u'' + (V-z) u = 0
\end{equation}
for $z\in \C$. To be more precise, define for differentiable $v$
on a subinterval $I\subset \R$, and $x\in I$, the vector
$\underline{v} (x)$ by $\underline{v} (x) = (v(x), v'(x))^t$,
where $t$ denotes the transpose. Then, for arbitrary $x,y\in \R$
and $z\in \C$, the transfer matrix is the unique $2\times
2$-matrix $M(x,y,z)$ with $M(x,y,z) \underline{u}(y) =
\underline{u} (x)$ for every solution $u$ of \eqref{homogen}.
$M(x,y,z)$ has columns $(u_N(x),u'_N(x))^t$ and
$(u_D(x),u'_D(x))^t$, where $u_N$ and $u_D$ are the solutions of
\eqref{homogen} which satisfy initial conditions $(1,0)^t$ and
$(0,1)^t$ at $x$, respectively.

As usual, the Wronski determinant $W(u_1,u_2)$ of solutions $u_1$, $u_2$ of
\eqref{homogen} is defined by $W (u_1,u_2) = u_1(x) u_2'(x) - u_1'(x) u_2 (x)$ and this
expression does not depend on $x$. This implies that $\det M(x,y,z)=1$.

Let $\ueins$ be the solution of \eqref{homogen} with $\uueins (0)=(1,0)^t$, $\unull$ the
solution of \eqref{homogen} with $\uunull (0) = (0,1)^t$.  For $ z\in \C$ with positive
imaginary part, let $\uinf$ be the solution of \eqref{homogen}, which is square
integrable (at $\infty$) and satisfies $\uinf(0) = 1$.

Finally, for a  measurable locally bounded $g$  and $f\in L^2
(\R)$ with compact support we write
$$
\langle g,f \rangle :=\int_\R g(t) \overline{f(t)} dt.
$$
In the situation we have in mind, $g$ will be a solution of
\eqref{homogen}.

\medskip

We will first state our result for the half line. For $\alpha>0$, $C>0$, and $N > 1$, we
define
$$
P(\alpha,C,N):=\{ E\in \R : \|M(x,y,E)\|\leq C N^\alpha \text{ for all } 0 \le x,y \le N
\}.
$$

We can now give a precise  version of   our main theorem.

\begin{theorem}\label{mainhl}
Suppose $E_0\in \R$ is such that there exist $C>0$ and $\alpha
>0$ with $E_0\in P(\alpha,C,N)$ for all sufficiently large $N$. Let
$A(N)$ be a subset of $P(\alpha,C,N)$ containing $E_0$ such that $\mathrm{diam}
(A(N))\longrightarrow 0$ as $N \to \infty$. Then, for every compactly supported $f\in L^2
(0,\infty)$ with $\langle u_{0,E_0}, f\rangle \neq 0$, there exists $\widetilde{C}>0$
such that for $T$ large enough,
$$
M_{f,D} (T,p)  \geq \widetilde{C}  |B(T)| T^{\frac{p - 3\alpha}{1
+ \alpha}},
$$
where $B(T)$ is the $1/T$ neighborhood of $A(T^{\frac{1}{1+\alpha}})$.
\end{theorem}

We will now  state our result for the whole line.  In this case, for $\alpha>0$, $C>0$,
and $N>1$, we define
$$
P(\alpha,C,N):=\{ E\in \R : \|M(x,y,E)\|\leq C N^\alpha \text{ for all } -N \le x, y \le
N \}.
$$
\begin{theorem}\label{mainwl}
Suppose $E_0\in \R$ is such that there exist $C>0$ and $\alpha >0$ with $E_0\in
P(\alpha,C,N)$ for sufficiently large $N$. Let $A(N)$ be a subset of $P(\alpha,C,N)$
containing $E_0$ such that $\mathrm{diam} (A(N))\longrightarrow 0$ as $N \to \infty$. Let
$f\in L^2 (\R)$  be compactly supported and satisfy $\langle u, f\rangle \neq 0$ for at
least one solution $u$ of \eqref{homogen} with $z = E_0$. Then, there exists
$\widetilde{C}>0$ such that for  $T$ large enough,
$$
M_f (T,p)  \geq \widetilde{C}  |B(T)| T^{\frac{p - 3\alpha}{1 + \alpha}},
$$
where $B(T)$ is the $1/T$ neighborhood of $A(T^{\frac{1}{1+\alpha}})$.
\end{theorem}

As in \cite{DT}, the previous theorems have the following
immediate consequences. We only state them for the half-line case.
They hold with obvious modifications for the whole line.

\begin{coro}\label{oneenergy}
Suppose there is an energy $E_0 \in \R$ such that $\|
M(x,y,E_0)\|\leq C N^\alpha$ for all $N$ large enough and $0 \le
x,y \le N$. Then, for every compactly supported $f\in L^2
(0,\infty)$ with $\langle u_{0,E_0}, f\rangle \neq 0$, we have
$$
\beta_{f,D}^-(p) \geq \frac{p - 1 - 4\alpha}{1 + \alpha},
$$
\end{coro}

\begin{proof}
We only need to take $A(N) = \{ E_0 \}$ for every $N$. Then $|B(T)| = 2T^{-1}$ and the
assertion follows.
\end{proof}

\begin{coro}\label{squareroot}
Suppose that there exist $C > 0$, $E_0 \in \R$, $0<\theta \le 1$
such that $\| M(x,y,E) \| \le C$ for all $N$, $0\le x,y \le N$ and
$E\in [E_0-N^{-\theta}, E_0 + N^{-\theta}]$. Then, for every
compactly supported $f\in L^2 (0,\infty)$ with $\langle u_{0,E_0},
f\rangle \neq 0$, we have
$$
\beta_{f,D}^-(p) \geq p - \theta.
$$
\end{coro}

\begin{proof}
Let $A(N) = [ E_0 - N^{-\theta} , E_0 + N^{-\theta} ]$. Then
$|B(T)| \ge |A(T)| = 2T^{-\theta}$ and the assertion follows.
\end{proof}

Since compactly supported perturbations of $V$ (and even perturbations with a suitable
power-decay; compare \cite{DST}) leave the power-law bounds of the form above unchanged,
these results immediately extend to all these perturbed models, whenever they apply. We
refer the reader to \cite{DST,DT} for a more detailed discussion of stability issues.

The proofs of Theorem~\ref{mainhl} and Theorem \ref{mainwl} will be given at the end of
this section. We first gather
a series of preliminary results that we will need in the proof.

\begin{lemma}\label{Kato}
Let $\mathcal{H}$ be a separable Hilbert space, $S$ a selfadjoint operator on
$\mathcal{H}$ and $A$ a closed operator on $\mathcal{H}$.  Then,
$$
2 \pi \int_0^\infty \exp \left( -\tfrac{2 t}{T} \right) \left\| \, A  \exp(-i t S) f
\right\|^2 dt = \int_{\R} \left\| \, A (S - E - \tfrac{i}{T})^{-1}  f \right\|^2 d E
$$
for every $f\in \mathcal{H}$ and $T > 0$.
\end{lemma}

\begin{proof}
This identity is well known. For example, it falls well within the discussion in
\cite[pp.~142--144]{rs4}. For the convenience of the reader we give a sketch of the
proof:

Let $T>0$ be given. Define $\varphi : \R \longrightarrow \mathcal{H}$ by
\begin{equation*}
\varphi(t) :=\left\{\begin{array}{r@{\quad:\quad}l}
   \exp( - \frac{t}{T}) \exp (- i t S) f& t >0,\\
0\in \mathcal{H}  & t \leq 0,
\end{array}\right.
\end{equation*}
and $\widehat{\varphi}  : \R  \longrightarrow \mathcal{H}$ by
$ \widehat{\varphi} (E) := \int_\R \exp (- i t E) \varphi (t) dt$.
Then,
$$
\exp \left( -\tfrac{2 t}{T} \right) \left\| \, A
\exp(-i t S) f \right\|^2 = \| A\varphi (t)\|^2
$$
and
$$
\widehat{\varphi} (E) = \int_0^\infty \exp (- i t E) \exp( - \tfrac{t}{T}) \exp (- i t S)
f dt = i (S + E - \tfrac{i}{T})^{-1} f.
$$
Now, (32) on \cite[p.~143]{rs4} says $2 \pi \int_\R \| A \varphi (t) \|^2 dt = \int_\R \|
A \widehat{\varphi} (E)\|^2 dE$ and the desired equality follows.
\end{proof}

\begin{lemma} \label{Sobolev}
For each $M\in (0,\infty)$, there is $C=C(M)<\infty$ such that for every $q:\R \to\C$
with $\|q\|_{1,\mbox{\rm\footnotesize unif}} \le M$ and each solution $u$ of $-u''+qu=0$,
\begin{equation} \label{eqsobolev}
|u' (x)|^2 \leq C \int_{x-1}^{x +1} |u(s)|^2 ds \quad \mbox{for
every $x\in \R$},
\end{equation}
and
\begin{equation} \label{eqsobcon}
\int_{a-1}^{a+1} \left(|u(x)|^2 +|u'(x)|^2 \right)\,dx \le (1+2C)
\int_{a-2}^{a+2} |u(x)|^2\,dx \quad \mbox{for every $a\in\R$}.
\end{equation}
\end{lemma}

\begin{proof}
The first statement is well known, see for example Lemma~A.3 in
\cite{DSS}. It implies that
\begin{eqnarray*}
\int_{a-1}^{a+1} |u' (x)|^2 dx &\leq & C \int_{a-1}^{a+1} \int_{x-1}^{x+1} |u (t)|^2 dt\,
dx\\ &\leq & C \int_{a-1}^{a+1} \int_{a-2}^{a+2} |u (t)|^{2} dt \,dx\\ &=& 2 C
\int_{a-2}^{a+2} |u (t) |^2 dt,
\end{eqnarray*}
which gives \eqref{eqsobcon}.
\end{proof}

\begin{lemma} \label{Gronwall}
Let $E \in \R$ and $N \ge 0$ be given. Define
$$
L(N):=\sup_{0 \le x,y \le N} \| M(x,y,E)\|.
$$
Then, for every $\delta \in \C$ and $0 \le x,y \le N$, we have
\begin{equation}\label{gw1}
\|M(x,y,E + \delta)\|\leq L(N) \exp \left( L(N) |x - y| |\delta|
\right).
\end{equation}
In particular, if $\| M(x,y,E)\|\leq C N^\alpha$ for all $0 \le
x,y \le N$, then
\begin{equation}\label{gw2}
\| M(x,y,E + \delta)\| \leq C \exp (C) N^\alpha
\end{equation}
whenever $0 \le x,y \le N$ and $0 \le |\delta| \le N^{-1-\alpha}$.
\end{lemma}

\begin{proof}
Essentially, \eqref{gw1} is \cite[Eq.~(3.2)]{Sim}. The estimate \eqref{gw2} is an
immediate consequence of \eqref{gw1}.
\end{proof}

The following lemma is the crucial new ingredient in our treatment
of the half-line operator.
\begin{lemma}\label{crucial}
For $z\in \C \setminus \R$, define $u_{f,z} = (H_D-z)^{-1} f$. Suppose $E \in \R$ and
$f\in L^2(0,\infty)$ with ${\rm supp} \,f \subset [0,s]$ are such that
\begin{equation} \label{infimum} 0 = \lim_{\delta \to 0 + }\inf\{ \|\uuf (s)\| : z \in \Cplus,
\, |z- E| \le \delta\}.
\end{equation}
Then, $ 0 = \langle u_{0,E}, f\rangle$.
\end{lemma}

\begin{proof}
By \eqref{infimum}, there exists a sequence $(z_n) $ in $\Cplus$ with $z_n \to E$ and
$\underline{u}_{f,z_n} (s) \to (0,0)^t$ for $n \to \infty$. By $u_{f,z_n} (0) = 0$ for
all $n$ and continuity, the inhomogeneous equation
$$
- u'' + (V - E) u = f
$$
has a solution $v$ with $v(0) = v(s) = v'(s) = 0$.
Let $Y (t)$ be the fundamental matrix of the homogeneous equation at $x=s$ (i.e., the
columns of $Y$, $(v_1, v_1')^t $ and $(v_2, v_2)^t$ are solutions $v_1,v_2$ of the
homogeneous equation which satisfy $\underline{v}_1 (s) = (1,0)^t$ and $\underline{v}_2
(s) = (0,1)^t$). Then,
$$
\underline{v} (x) = Y(x) \int_s^x Y(t)^{-1} (0,f(t))^t dt.
$$
Restricting our attention to the first component, we obtain
\begin{equation}\label{ufprod}
0 = v(0) = \int_s^0 [ - v_1(0) v_2 (t) + v_2 (0) v_1 (t)] f(t) dt.
\end{equation}
Now, obviously,
$$
u (t) : = [ - v_1(0) v_2 (t) + v_2 (0) v_1 (t)]
$$ is a solution of the homogenous equation
with $u(0) = 0$. As $v_1$
and $v_2$ are linearly independent, $u$ does not vanish identically. Thus, $u$ agrees up
to a non-vanishing factor with $ u_{0,E}$. The assertion of the lemma therefore follows
from \eqref{ufprod}.
\end{proof}

To treat the whole line operator we will use a variant of the
lemma. It is given as follows.

\begin{lemma}\label{crucialwl}
For $z\in \C \setminus \R$, define $\uuf = (H-z)^{-1} f$. Let $E\in \R$ and $f\in
L^2(\R)$ with ${\rm supp} \,f\subset [-s,s]$ be such that
\begin{equation} \label{infimumwl}
0 = \lim_{\delta \to 0 + }\inf\{ \|\uuf (s)\| + \|\uuf(-s)\| : z \in \Cplus, \, |z- E|
\le \delta\}.
\end{equation}
Then, $\langle u , f\rangle =0 $ for every solution $u$ of $ - u'' + V u = E u$.
\end{lemma}

\begin{proof}
By \eqref{infimumwl}, there exists a sequence $(z_n)$ in $\Cplus$ with
$z_n \to E$ and $\underline{u}_{f,z_n} (s) \to (0,0)^t$ and $\underline{u}_{f,z_n} (-s)
\to (0,0)^t$ for $n \to \infty$.  Let $v$ be the solution of $ - v'' + (V - E) = f$ with
$\underline{v} (s) = (0,0)^t$. Then, by continuous dependence of solutions on initial
conditions, $\underline{u}_{f,z_n} (x)\longrightarrow \underline{v}(x)$ for every $x\in
\R$.  In particular, $\underline{v} (-s) = (0,0)^t$. Let $Y(t)$ be as in the proof of the
previous lemma. Then,
$$
\underline{v} (x) = Y(x) \int_s^x Y(t)^{-1} (0,f(t))^t dt =
\int_{s}^x\left( \begin{array}{c} - v_1 (x) v_2 (t) + v_2 (x) v_1 (t)
\\ - v_1' (x) v_2 (t) + v_2' (x) v_1 (t) \end{array} \right) f(t) dt.
$$
Thus,
$$\left( \begin{array}{c} 0 \\ 0 \end{array} \right) = \underline{v}(-s) = \int_{-s}^s
\left( \begin{array}{c} v_1 (-s) v_2 (t) - v_2
(- s) v_1 (t) \\ v_1' ( -s ) v_2 (t) - v_2' (- s ) v_1 (t) \end{array} \right) f(t) dt.
$$

Now, $u_1 := v_1 (-s) v_2 - v_2 (- s) v_1$ and $u_2 := v_1' (-s ) v_2 - v_2' (- s) v_1 $
are solutions of $- u'' + V u = Eu$. They satisfy $\underline{u}_1 (-s) =
(0,W(v_1,v_2))^t = (0,1)$ and $\underline{u}_2 (-s)  = (-W(v_1,v_2),0)^t= (-1,0)^t$.
Thus, $u_1$, $u_2$, are a fundamental system for $- u'' + V u = Eu$ and we have shown
that $\langle u_i, f\rangle =0$, $i=1,2$. This completes the proof of the lemma.
\end{proof}

\begin{proof}[Proof of Theorem~\ref{mainhl}.]
As before we set $u_{f,z} := (H_D -  z )^{-1} f$. Let $s>0$ with
supp$\,f \subset [0,s]$ and define $N(T) :=
T^{\frac{1}{1+\alpha}}$.

Now, we can apply Lemma~\ref{Kato} with $S=H_D$ and $A =
|\cdot|^{\frac{p}{2}}$ and Lemma~\ref{Sobolev} with $u=u_{f,z}$.

For $T$ large enough, this gives
\begin{eqnarray} \label{longcalc}
M_{f,D} (T,p) &=& \tfrac{2}{T} \int_0^\infty \exp \left( -\tfrac{2
t}{T} \right) \left\| \, |\cdot|^{p/2} \exp(-i t H_D) f \right\|^2
dt \nonumber \\
(\mbox{Lemma \ref{Kato}})\:\;&=& \int_{\R} |x|^p \tfrac{1}{\pi T}
\int_{\R} |u_{f, E + \frac{i}{T}} (x) |^2 d E\, dx \nonumber \\
&\geq & \frac{1}{4} \sum_{n=s + 2}^\infty (n-2)^p \int_{n-2}^{n+2}
\tfrac{1}{\pi T} \int_{\R} |u_{f, E + \frac{i}{T}} (x) |^2 d E\,
dx\nonumber \\
&\geq& \frac{1}{4} \sum_{n=s +
  2}^\infty (n-2)^p \int_{ B(T) } \int_{n-2}^{n+2} \tfrac{1}{\pi T}
|u_{f, E + \frac{i}{T}} (x) |^2 d x \, d E \nonumber \\
&\geq& \frac{c}{T} \sum_{n=s +
  2}^\infty (n-2)^p \int_{ B(T) } \int_{n-1}^{n+1}
\|\underline{u}_{f, E + \frac{i}{T}} (x) \|^2 d x \, d E.
\end{eqnarray}
In the last step, Lemma~\ref{Sobolev} was used, based on the fact that $u_{f,E+i/T}$ is a
solution of $-u''+Vu= (E+i/T)u$ on $[n-2,n+2]$. Observe that the constant $c>0$ can be
chosen uniformly for all sufficiently large $T$. Using that the transfer matrices satisfy
$\|M^{-1}\|=\|M\|$, we can further bound \eqref{longcalc} from below by
\begin{eqnarray*}
& \geq & \tfrac{c}{T} \sum_{n=s + 2}^\infty (n-2)^p \int_{B(T)}
\int_{n-1}^{n+1} \|M(x,s, E +\tfrac{i}{T})\|^{-2}
\|\underline{u}_{f, E + \tfrac{i}{T}} (s)\|^2 dx\, d E\\
&\geq & \tfrac{c}{T} \sum_{n=\frac{N(T)}{2} + 2}^{N(T)-1} (n-2)^p
\int_{B(T) } \int_{n-1}^{n+1} \|M(x,s, E +\tfrac{i}{T})\|^{-2}
\|\underline{u}_{f, E + \tfrac{i}{T}} (s)\|^2 dx \, dE\\
\;\:&\geq & \tfrac{2c}{T}
\sum_{n=\frac{N(T)}{2} + 2}^{N(T)-1} \left(\tfrac{N(T)}{2}\right)^p
\int_{B(T)} (C \exp (C) N(T)^{\alpha})^{-2} \, \|\underline{u}_{f, E +
  \tfrac{i}{T}} (s)\|^2 dE\\
  &\ge & \tfrac{2^{1-p} c}{T}
\tfrac{N(T)}{3} N(T)^p |B(T)| (C \exp (C) N(T)^{\alpha})^{-2}
\inf_{\mbox{\tiny dist}(z,B(T)) \le \tfrac{1}{T}}
\|\underline{u}_{f,z} (s) \|^2.
\end{eqnarray*}
Here, we used Lemma~\ref{Gronwall} in the second to the last step.

By Lemma \ref{crucial} and $\langle u_{0,E}, f\rangle \neq 0$, there
exists $ \kappa >0$ and $\delta >0$ with $\inf\{ \| \uuf (s)\|^2 : |z
- E_0|\leq \delta \} \geq \kappa $. By ${\rm diam}(A(N))\to 0$ as $N
\to \infty$ and $E_0\in A(N)$ for all $N$, we obtain
$$ \inf \{ \| \uuf (s) \|^2 : \mbox{dist}(z,B(T))\leq \tfrac{1}{T} \}
\geq \kappa >0
$$ for $T$ sufficiently large.

Thus, we can summarize the above estimates as
$$
M_{f,D} (T,p)  \geq \widetilde{C}  |B(T)| T^{ \frac{p - 3 \alpha}{
1 + \alpha}  }.
$$
This proves the theorem.
\end{proof}

\begin{proof}[Proof of Theorem~\ref{mainwl}.]
We set $u_{f,z} := (H - z )^{-1} f$.  By Lemma \ref{crucialwl}, there
exists $\epsilon \in \{-1,1\}$, $\kappa >0$ and $\delta>0$ such that
$$ \inf \{ \| \uuf (\epsilon s) \|^2 : |z - E_0|\leq \delta\} \geq
\kappa.$$ Now, we consider the positive half-line if $\epsilon=1$ and
the negative half-line if $\epsilon =-1$. The proof is then a simple
modification of the argument used in the proof of Theorem~
\ref{mainhl}.
\end{proof}

\section{The Bernoulli-Anderson model: Basic setting and deterministic results}
\label{deterministic}

In the next three sections we consider the following situation: Let $g_0$ and $g_1$ be
two real-valued, locally integrable potentials with support in $[0,1]$. Also, let $\omega
= (\omega_n)_{n\in\Z}$ be a two sided sequence with $\omega_n \in \{0,1\}$ for all $n$.
Define the Schr\"odinger operator
\begin{equation} \label{eq4.1a}
H_{\omega} = -\frac{d^2}{dx^2} + V_{\omega},
\end{equation}
where
\begin{equation} \label{eq4.1b}
V_{\omega}(x) = \sum_{n\in\Z} g_{\omega_n}(x-n)
\end{equation}
in $L^2(\R)$.
We may
equivalently write
\begin{equation} \label{eq4.2}
H_{\omega} = -\frac{d^2}{dx^2} + V_{per}^{(0)}(x) + \sum_{n\in\Z} \omega_n g(x-n),
\end{equation}
with deterministic periodic background potential $V_{per}^{(0)}(x)
= \sum_n g_0(x-n)$ and single-site potential $g=g_1-g_0$. In the
case where the $\omega_n$ are independent, identically distributed
random variables, the family $H_{\omega}$ then represents a
continuum Bernoulli-Anderson-type model. This case will be
considered in Section~\ref{almostsure}, while we state a simple
deterministic bound in this section.

Define also $V_{per}^{(1)}(x) = \sum_n g_1(x-n)$ and consider the
two periodic Schr\"odinger operators $H^{(j)} = -d^2/dx^2 +
V_{per}^{(j)}$. Let $T_j(E)$ be the transfer matrix for $H^{(j)}$
at energy $E$ from $0$ to $1$.

We say that $E_0 \in \R$ is a critical energy for $H_{\omega}$ if
\begin{equation} \label{eq4.3}
\begin{array}{l} \mbox{(i) $T_0(E_0)$ and $T_1(E_0)$ commute}, \\ \mbox{(ii)
$E_0$ is contained in the interior of the spectra of $H^{(0)}$ and
of $H^{(1)}$.} \end{array}
\end{equation}

The same definition was used for discrete polymer models in
\cite{JSBS}. With the help of Theorem~\ref{mainwl} we can now
extend the results obtained in \cite{JSBS} to continuum operators,
starting with a continuum analog of Theorem~1 in \cite{JSBS}.

\begin{lemma} \label{lem4.1}
If $E_0$ is a critical energy for $H_{\omega}$, then the transfer
matrix of $H_{\omega}$ at $E_0$ is globally bounded: There exists
$C<\infty$ such that
\begin{equation} \label{eq4.4}
\| M(x,y,E_0)\| \le C
\end{equation}
for all $x,y \in \R$.
\end{lemma}

By the whole-line version of Corollary~\ref{oneenergy} (with $\alpha=0$), this has the
following immediate consequence.

\begin{theorem} \label{thm4.2}
If $f \in L^2(-s,s)$ is not orthogonal in $L^2(-s,s)$ to the space of solutions of
$-u''+V_{\omega}u = E_0u$, then
\begin{equation} \label{eq4.5}
\beta_f^-(p) \ge p-1.
\end{equation}
\end{theorem}

\begin{proof}[Proof of Lemma~\ref{lem4.1}.] By \eqref{eq4.3}(ii), the two transfer matrices
$T_j(E_0)$ each have two different complex-conjugate eigenvalues
$e^{\pm i \eta_j} \not\in \{\pm 1\}$ or are $\pm I$ (in which case
we set $\eta_j =0$ or $\eta_j = \pi$). Due to commutation, there
exists a real invertible matrix $M$ such that
\begin{equation} \label{eq4.6}
M T_j(E_0) M^{-1} = \left( \begin{array}{cr} \cos \eta_j & -\sin \eta_j \\
\sin \eta_j & \cos \eta_j \end{array} \right)
\end{equation}
simultaneously for $j=0$ and $j=1$. This shows that the transfer matrix of $H_{\omega}$
at $E_0$ between two given integers $x$ and $y$ is similar (via $M$) to a product of
rotations, and thus has norm bounded by $C = \|M\| \|M^{-1}\|$. Standard arguments (e.g.,
\cite[Appendix~A]{DSS}) imply that (\ref{eq4.4}) holds for arbitrary $x,y \in \R$ and
suitably enlarged $C$.
\end{proof}

Note that Lemma~\ref{lem4.1} is an entirely deterministic result:
If a critical energy $E_0$ exists (which only depends on $g_0$ and
$g_1$), then (\ref{eq4.4}) holds for {\it every} choice of the
sequence $\omega$. Similarly, the dependence on $\omega$ enters
Theorem~\ref{thm4.2} only through the non-orthogonality condition
on $f$, and thus involves only finitely many $\omega_n$,
($n=-s,\ldots,s-1$ if $s$ is an integer). The bound (\ref{eq4.5})
then holds uniformly in the values of all other $\omega_n$.

\section{The Bernoulli-Anderson model: Almost sure results}
\label{almostsure}

We now consider the model (\ref{eq4.1a}), (\ref{eq4.1b}) for
independent, identically distributed Bernoulli random variables
$\omega_n$, i.e.\ we equip $\Omega := \{0,1\}^{\Z}$ with the
measure $P = \prod_{j\in\Z} \mu$, where $\mu$ is a Bernoulli
probability measure on $\{0,1\}$, $\mu(\{0\})=p$, $\mu(\{1\})=1-p$
for some $0<p<1$.

For this case, under a slight restriction on the phases $\eta_0$
and $\eta_1$ from the proof of Lemma~\ref{lem4.1}, we will improve
the result from the previous section and show that the lower
diffusion exponents almost surely satisfy the lower bound $p-1/2$.
This is a continuum analog of Theorem~4 in \cite{JSBS}, which
establishes the same almost sure lower bound for discrete random
polymer models.

This will be achieved by combining Corollary~\ref{squareroot} with
a Borel-Cantelli argument and using a large deviations analysis of
the growth of transfer matrices for energies near a critical
energy. Here we follow the ideas developed for discrete models in
\cite{JSBS}.
\medskip

Our aim is to analyze the growth behavior of the transfer matrices
$$ T_\omega (k,m,E) := T_{\omega_{k-1}} (E) \ldots T_{\omega_m} (E)$$
for $\omega \in \{0,1\}^{\Z}$ and $E $ close to $E_c$. This can very conveniently be done
by a Pr\"ufer type decomposition, i.e.\ by decomposing the action of the transfer
matrices into a rotation and a scaling.

To do this simultaneously for all $E$ close to $E_c$, we need the following lemma.

\begin{lemma}\label{periodic}
Let $E_c$ be a critical energy. Then, there exists an interval $I$ around $E_c$,
$\eta_j\in \R$, and analytic functions $a_j, b_j : I \longrightarrow \C$, $j=0,1$, $F : I
\longrightarrow \mathrm{GL}(2,\C)$ such that
\begin{equation} \label{eqscattmatrix}
\Tjt (E) := F(E)^{-1} T_j (E) F(E) = \left( \begin{array}{cc} a_j (E) & \overline{b_j
(E)} \\ b_j (E) & \overline{a_j (E)} \end{array} \right) \quad \text{ for } E\in I
\end{equation}
with
$$b_j (E_c) =0,\;\:\mbox{and}\;\: a_j (E_c) = e^{i\eta_j} \;\:\mbox{for $\eta_j\in[0,2\pi)$}, \;\: j=0,1.$$
In fact, we may choose $b_j(E)=0$ and $|a_j(E)|=1$ for all $E\in
I$ and either $j=0$ or $j=1$. Moreover, $1 = \det \Tjt (E) = |a_j
(E)|^2 - |b_j (E)|^2$.
\end{lemma}

\begin{proof}
As $E_c$ is in the interior of both periodic spectra, we have
$$
D_j(E_c) := \mbox{tr}\,T_j(E_c) \in [-2,2]
$$
for $j=1,2$. W.l.o.g.\ we may assume that either $D_0(E_c)\in (-2,2)$ or that
$|D_j(E_c)|=2$ for both values of $j$. In the latter case $E_c$ is a degenerate gap for
$H^{(0)}$ and $H^{(1)}$ and thus $T_0(E_c)$ and $T_1(E_c)$ are both either $I$ or $-I$.

By Lemma~\ref{analyticev}, there is an open neighborhood $I$ of
$E_c$ and complex conjugate analytic $v_{\pm}(E)$ which for each
$E\in I$ are linearly independent eigenvectors of $T_0(E)$ to
complex conjugate analytic eigenvalues $\rho_{\pm}(E)$.

The matrix $F(E) :=(v_+ (E), v_- (E)) $ is invertible and
$$
F(E)^{-1} T_0 (E) F(E) = \left( \begin{array}{cc} \rho_+ (E) & 0
\\ 0 & \overline{\rho_+ (E)} \end{array} \right).
$$
Thus \eqref{eqscattmatrix} holds for $j=0$ with $b_0(E)=0$ and $a_0(E)= \rho_+(E)$ for
all $E\in I$. Moreover, as $T_1 (E)$ is real and the columns of $F(E)$ are complex
conjugates of each other, there exists $a_1$ and $b_1$ with
$$
F(E)^{-1} T_1 (E) F(E) = \left( \begin{array}{cc} a_1 (E) &
\overline{b_1 (E)} \\ b_1 (E) & \overline{a_1 (E)} \end{array}
\right). $$
As $F$ and $T_j$ are analytic, so are $a_j$ and $b_j$.
As $T_j$ has determinant equal to one by constancy of the
Wronskian, we have
$$
1 = \det \Tjt (E) = |a_j (E) |^2 - |b_j (E)|^2.
$$
Finally, the linearly independent eigenvectors $v_+(E_c)$ and $v_-(E_c)$ of $T_0 (E_c)$
are eigenvectors of $T_1 (E_c)$ as well (this is trivial if $|D_0(E_c)|=|D_1(E_c)|=2$ and
in the other case follows from the fact that $T_0(E_c)$ and $T_1(E_c)$ commute and that
$T_0(E_c)$ has one-dimensional eigenspaces). We infer $b_1 (E_c) = 0$ and $ |a_1 (E_c)| =
1$.
\end{proof}

Given this lemma, we can give the Pr\"ufer type analysis of the action of the transfer
matrices mentioned above. This will be carried out on the level of the $\Tjt$.  We
identify $\Seins$ with $[0,2\pi)$. For $\theta\in \Seins$, we define the unit vector
$\ettheta$ by
$$
\ettheta:= \frac{1}{\sqrt{2}}  \left( \begin{array}{c} \exp(i\theta)\\
\exp(- i \theta) \end{array} \right).
$$
Then,
$$
\Tjt (E) \ettheta =  \frac{1}{\sqrt{2}}
\left( \begin{array}{c} a_j (E)  \exp(i\theta) + \overline{b_j (E)}
\exp(- i \theta) \\
b_j (E)    \exp(i\theta) + \overline{a_j (E)} \exp(-i \theta)
\end{array} \right).
$$
Obviously, the first and the second component of $\Tjt (E) \ettheta$ are complex conjugates
of each other. Thus, for each $\delta:= E- E_c$, there exists a unique map
$$\Sdeltaj : \Seins \longrightarrow \Seins$$
with
$$ \Tjt (E) \ettheta = \|\Tjt (E) \ettheta \| \widetilde{e}_{\Sdeltaj (\theta)}$$
for all $\theta \in \Seins$.  Moreover, by $|a_j|^2 - |b_j|^2 =1$,
we find
\begin{equation}\label{norm}
\|\Tjt (E) \ettheta\|^2 = 1 + 2 \Re (a_j (E) b_j (E) e^{2 i \theta})
+ 2 |b_j (E)|^2.
\end{equation}

In order to study the transfer matrices it will be convenient to define iterates of the
$\Sdeltaj$. More precisely, for $l,m\in \Z$ with $l\geq m$ we define inductively
$$
\mathcal{S}_{\delta,\omega}^{m,m} (\theta)= \theta, \;\:\mathcal{S}_{\delta,\omega}^{l+1,m}
(\theta) = \mathcal{S}_{\delta,\omega_l} ( \mathcal{S}_{\delta,\omega}^{l,m} (\theta)
).
$$

\begin{prop}\label{normviatheta}
Let $M$ be a matrix of the form $
\left( \begin{array}{cc} a  & \overline{b} \\ b & \overline{a} \end{array}
\right)$. Then,
$$\|M\|= \sup_{\theta\in \Seins} \|M \ettheta\|. $$
\end{prop}
\begin{proof}
Let $Q:= \frac{1}{\sqrt{2}} \left( \begin{array}{cr} 1 & i \\ 1 & -i
\end{array} \right)$. As $Q$ is unitary, we have $\|M\| = \| Q^{-1} M
Q\|$. By assumption on $M$, the matrix  $Q^{-1}M Q$ is  real.
Thus, setting $e_\theta:= (\cos(\theta), \sin(\theta))^t$ and
using $\ettheta= Q e_\theta$, we obtain

\begin{equation*} \|Q^{-1} M Q\| = \sup_{\theta} \|Q^{-1} M Q e_\theta\|= \sup_{\theta} \| Q^{-1}
M \ettheta\| = \sup_{\theta} \| M \ettheta\|.
\end{equation*}
This finishes the proof.
\end{proof}

We are now in a position to provide the key expression for the norm of the transfer
matrix.

\begin{prop} \label{expressionnorm}
Let $\omega\in \{0,1\}^{\Z}$, $\delta\in \R$, $k,m\in \Z$ with $k>
m$ be given. Then,
$$\log\|T_\omega (k,m,E_c+ \delta)\|^2 = 2 \delta \sup_\theta \{\Re
\sum_{l=m}^{k-1} c_{\omega_l} e^{2 i
\mathcal{S}_{\delta,\omega}^{l,m} (\theta)} \} + O(\delta^2 (k-m),
1),$$ where $c_{\omega_l} :=e^{i\eta_{\omega_{\ell}}} \frac{d
b_{\omega_l} }{d E} (E_c)$.
\end{prop}

\begin{proof} Let $E = E_c + \delta$. We begin by estimating $\| F(E)^{-1} T_\omega (k,m,E) F(E)\|^2$:
\begin{align*}
\| F(E)^{-1} T_\omega (k,m,E) F(E)\|^2 &= \|\widetilde{T}_{\omega_{k-1}} (E) \ldots
\widetilde{T}_{\omega_{m}} (E) \|^2\\ (\mbox{Prop \ref{normviatheta}})\;\:& =
\sup_{\theta} \|\widetilde{T}_{\omega_{k-1}} (E) \ldots \widetilde{T}_{\omega_{m}} (E)
\ettheta\|^2 \\ &= \sup_{\theta} \prod_{l=m}^{k-1} \|\widetilde{T}_{\omega_l} (E)
\widetilde{e}_{\mathcal{S}_{\delta,\omega}^{l,m} (\theta)}\|^2\\
\eqref{norm}\;\:&= \sup_{\theta} \prod_{l=m}^{k-1} ( 1 + 2\Re (a_{\omega_l}(E)
b_{\omega_l} (E) e^{2 i \mathcal{S}_{\delta,\omega}^{l,m} (\theta)}) +  2 |b_{\omega_l}
(E)|^2).
\end{align*}
As $b$ is analytic around $E_c$ with $b_j(E_c)=0$ for $j=0,1$, by
Lemma \ref{periodic}, we have $b(E_c + \delta) = O(\delta)$. Thus,
taking logarithms and invoking $\log (1 + x) = x + O(x^2)$, we obtain
from the previous formula
$$\log\| F(E)^{-1} T_\omega (k,m,E) F(E)\|^2 = 2 \sup_{\theta} \Re
\sum_{l=m}^{k-1} a_{\omega_l}(E) b_{\omega_l}(E) e^{2 i
\mathcal{S}_{\delta,\omega}^{l,m} (\theta)} + O(\delta^2 (k-m)).$$
By analyticity of   $b_j$ and $a_j$ around $E_c$ we further have
$b_j (E) = \delta \frac{d b_j }{d E} (E_c) + O(\delta^2)$ and
$a_j(E) = a_j(E_0)+O(\delta) = e^{i\eta_j} +O(\delta)$. Thus, we
end up with
$$\log\| F(E)^{-1} T_\omega (k,m,E) F(E)\|^2 = 2 \delta
\sup_{\theta} \Re \sum_{l=m}^{k-1} c_{\omega_l} e^{2 i
\mathcal{S}_{\delta,\omega}^{l,m} (\theta)} + O(\delta^2 (k-m)).$$
The statement of the proposition follows as $F(E)$ and its inverse
$F(E)^{-1}$ are uniformly bounded in a neighborhood of $E_c$.
\end{proof}

The proposition gives a closed expression for the norm of the
transfer matrices in terms of sums of the form
$$
\sum_{l=m}^{k-1} c_{\omega_l} e^{2 i \Smitallem (\theta)}.
$$

For sums of this form, a large deviation estimate has been proven in \cite{JSBS} in the
case where $\{0,1\}^{\Z}$ is equipped with a Bernoulli measure. This estimate carries
over to our situation almost immediately. Here are the details:

\begin{definition}
Set $\Smitallemn :=S_{\delta,\omega}^{k,0} $. Let $\alpha >0$,
$N\in \N, \omega \in\{0,1\}^{\Z}$, $\delta\in \R$ and $\theta\in
\Seins$ be given. Define $I_{\omega,N} (\theta,\delta)
:=\sum_{k=0}^{N-1} c_{\omega_k} e^{2 i \Smitallemn (\theta)}$ and
$$\Omega_N (\alpha,\delta,\theta) :=\{\omega\in \Omega : \exists k\leq
N \;\:\mbox{s.t.}\;\: |I_{\omega,k} (\theta,\delta)|\geq N^{\alpha +
\frac{1}{2}}\}. $$
\end{definition}

We consider $\Omega$ as a probability space with the Bernoulli
measure $P$ defined at the beginning of this section. As above
$\eta_0$ and $\eta_1$ are the phases of $a_0(E_c)$ and $a_1(E_c)$.

\begin{lemma} \label{largedeviation}
Assume that $\eta_0-\eta_1$ is not an integer multiple of $\pi$.
Then, for every $\alpha>0$, there exist $C_1<\infty$ and $C_2>0$
such that for all $\theta\in \Seins$ and all $\delta\in \R$ and
$N\in \N$ with $\delta^2 N \leq 1$, the estimate
$$
P (\Omega_N (\alpha,\delta,\theta)) \leq C_1 e^{- C_2 N^\alpha }
$$
holds.
\end{lemma}

\begin{proof}
Let $E= E_c + \delta$. By definition of the action $\mathcal{S}$ we have $\Tjt  (E)
\ettheta = \| \Tjt (E) \ettheta\| \widetilde{e}_{\Sdeltaj (\theta)}$. Combining this with
\eqref{norm}, we find
$$
\Tjt (E) \ettheta = \widetilde{e}_{\mathcal{S}_{\delta,j} (\theta)} + O(|b_j(E)|).
$$
On the other hand, the analyticity shown in Lemma \ref{periodic} gives
$$
\Tjt (E) \ettheta = \widetilde{e}_{\theta+ \eta_j} + O (\delta).
$$
Combining the last two equalities and using $b_j(E_c + \delta) = O(\delta)$, we obtain
$$
\widetilde{e}_{\Sdeltaj(\theta)} =   \widetilde{e}_{\theta+ \eta_j} + O(\delta),
$$
from which we conclude
$$
e^{2 i \Sdeltaj(\theta)} = e^{2 i (\theta + \eta_j)} + O (\delta).
$$
This formula is the crucial input in the proof of Theorem 6 of
\cite{JSBS}. Thus, we can now follow this proof line by line to
obtain the desired statement. We only note that the condition
$|pe^{2i\eta_0}+(1-p)e^{2i\eta_1}|<1$ used in this context in
\cite{JSBS} is equivalent to our condition $\eta_0 \not= \eta_1
\mod \pi$ (as $pe^{2i\eta_0}+(1-p)e^{2i\eta_1}$ is a
convex-combination of two numbers on the unit circle).
\end{proof}

We can now state our main result on Bernoulli-type models.

\begin{theorem}\label{dreiviertelschranke}
Assume that $\eta_0-\eta_1$ is not an integer multiple of $\pi$. Let $\alpha>0$ be
arbitrary. Then, there are $c>0$ and $C<\infty$ such that for every $N \in \N$, there is
a set $\Omega_N(\alpha)\subset \{0,1\}^\Z$ with $P(\Omega_N (\alpha)) \le Ce^{-c
N^{\alpha}}$ and
$$
\| T_\omega (x,y, E)\|\leq C
$$
for all $\omega \in \Omega \setminus \Omega_N(\alpha)$, $-N\leq x,
y\leq N$ and $|E - E_c|\leq N^{-\alpha - 1/2}$.

In particular, $\beta_f^-(p) \ge p-1/2$ holds for almost every
$\omega$ and every compactly supported $f$ that is not orthogonal
to all solutions of $-u''+V_{\omega}u=E_0 u$.
\end{theorem}

\begin{proof}
The first claim is established by following the proof of Theorem~6 in \cite{JSBS}: By
translation invariance it suffices to consider $0 \le x,y \le 2N$. Let
$\Omega_N(\alpha,\delta) := \Omega_{2N}(\alpha,\delta,0) \cap
\Omega_{2N}(\alpha,\delta,\pi/2)$. Then, by Lemma~\ref{largedeviation},
$P(\Omega_N(\alpha,\delta)) \le C_1' e^{-C_2' N^{\alpha}}$. From
Proposition~\ref{expressionnorm} it follows that there is a constant $C'<\infty$ such
that for all $N\in \N$, integers $0\le k,m \le 2N$, $|\delta|\le N^{-\alpha-1/2}$ and
$\omega \in \Omega_N(\alpha,\delta)$ it holds that
\begin{equation} \label{eqtransbound}
\|T_{\omega}(k,m,E_c+\delta) \| \le C'.
\end{equation}
Here we have also used that $T(k,m) = T(k,0) T(m,0)^{-1}$ and that
$\|A\| = \sup_{\theta} \|Ae_{\theta}\| \le \sqrt{2} \{\|A e_0\|,
\|A e_{\pi/2}\| \}$ for every $2\times 2$-matrix $A$. As remarked
at the end of the proof of Theorem~\ref{thm4.2}, the bound
(\ref{eqtransbound}) extends to transfer matrices between
arbitrary real $x,y\in [0,2\pi]$.

Note that so far $\delta$ is fixed in $\Omega_N(\alpha,\delta)$.
To find a set $\Omega_N(\alpha)$ such that transfer matrices for
$\omega \in \Omega_N(\alpha)^c$ are bounded uniformly for all
$|\delta| \le N^{-\alpha-1/2}$ we use Lemma~\ref{Gronwall}. Set
$\varepsilon = N^{-\alpha -1/2}$ and
$$
\Omega_N(\alpha) = \bigcup_{\ell=-N}^N \Omega_N(\alpha,
\ell\epsilon/N).
$$
For fixed $\ell$, Lemma~\ref{Gronwall} shows that $\|T_{\omega}(x,y,E_c+\delta)\|$ is
uniformly bounded for $\omega \in \Omega_N(\alpha,\ell\varepsilon/N)$, $\delta \in
[\ell\varepsilon/N - \frac{1}{N}, \ell\varepsilon/N + \frac{1}{N}]$ and $0\le x,y \le
2N$. This establishes the first part of the theorem as $P(\Omega_N(\alpha)) \le 2NC_1'
e^{-C_2'N^{\alpha}} \le Ce^{-cN^{\alpha}}$.

The lower bound on diffusion exponents now follows from (the whole-line version) of
Corollary~\ref{squareroot}. For each $\alpha>0$, $P(\Omega_N(\alpha))$ is summable over
$N$. Thus, by Borel-Cantelli, the assumption of Corollary~\ref{squareroot} is satisfied
for almost every $\omega$ and $\theta = \frac{1}{2} + \alpha$. We get that almost surely
$\beta_f^-(p) \ge p -\frac{1}{2} -\alpha$ for every compactly supported $f$ that is not
orthogonal to all solutions of $-u''+V_{\omega}u=E_0u$. We finally take $\alpha =
\frac{1}{n} \to 0$, using a countable intersection of full measure sets.
\end{proof}

\section{The Bernoulli-Anderson model: A concrete example}\label{sec5}

The existence of critical energies for a given pair $g_0$ and $g_1$ is not a generic
property. In fact, as (\ref{eq4.4}) immediately implies vanishing of the Lyapunov
exponent, the set of critical energies must be discrete by the results of \cite{DSS}.
However, it is easy to give examples where critical energies exist, see \cite{DSS2}. The
most simple one is given by $g_0=0$ and $g_1= \lambda \chi_{[0,1]}$, $\lambda>0$, that
is, $V_{\omega}$ consists of constant steps of height $0$ or $\lambda$.  This example
will be discussed in more detail in this section.

\medskip

In this case, all energies $E>\lambda$ satisfy \eqref{eq4.3}(ii). For such energies, the
transfer matrices are
\begin{equation} \label{eq4.7}
T^{(0)}(E) = \left( \begin{array}{cc} \cos k & \frac{1}{k} \sin k
\\ -k \sin k & \cos k \end{array} \right)
\end{equation}
and
\begin{equation} \label{eq4.8}
T^{(1)}(E) = \left( \begin{array}{cc} \cos \alpha & \frac{1}{\alpha} \sin \alpha
\\ -\alpha \sin \alpha & \cos \alpha \end{array} \right)
\end{equation}
where $k=\sqrt{E}$ and $\alpha = \sqrt{E-\lambda}$. If $E=n^2\pi^2$, $n\in\N$, then
$T^{(0)}(E) = \pm I$. On the other hand, if $E= n^2\pi^2 +\lambda$, then $T^{(1)}(E) =
\pm I$. In both cases, $T^{(0)}(E)$ and $T^{(1)}(E)$ commute. Thus, when $\lambda \in
(0,\pi^2)$, we have the following critical energies:
\begin{equation} \label{eq4.9}
\{ n^2 \pi^2: n\in \N\} \cup \{ n^2\pi^2 +\lambda: n\in\N\}.
\end{equation}

The richness of the set of critical energies gives us considerable flexibility in
choosing the initial state $f$ in Theorem~\ref{thm4.2} and
Theorem~\ref{dreiviertelschranke}, respectively. In fact, every non-zero $f$ with support
in $[0,1]$ satisfies the required non-orthogonality condition for at least one of the
critical energies and we get from Theorem~\ref{thm4.2}:

\begin{coro} \label{cor4.3}
Let $\lambda \in (0,\pi^2)$, $g_0=0$, $g_1 = \lambda \chi_{[0,1]}$, and $H_{\omega}$ be
given by \eqref{eq4.1a} and \eqref{eq4.1b}. Then
\begin{equation} \label{eq4.10}
\beta_f^-(p) \ge p-1
\end{equation}
for any $f\in L^2(0,1)$, $f\not= 0$ and any $\omega$.
\end{coro}

\begin{proof}
There is at least one $n\in \N$ such that
\begin{equation} \label{eq4.12}
\int_0^1 f(x) \sin(\pi nx) \,dx \not= 0 \quad \mbox{or} \quad
\int_0^1 f(x) \cos(\pi nx) \,dx \not= 0.
\end{equation}
If $f \not= \mbox{const}$ on $[0,1]$, then this follows (with even $n$) as $\{1\} \cup \{
\sin(2\pi kx), \cos(2\pi kx); k\in \N\}$ span $L^2(0,1)$. For $0\not= f = \mbox{const}$,
we may choose $n=1$.

Now consider two cases: If $\omega_0 = 0$, that is, $V_{\omega}(x)=0$ on $[0,1]$, then by
\eqref{eq4.12}, $f$ is not orthogonal in $L^2(0,1)$ to the space of solutions of $-u''=
n^2\pi^2 u$. Thus, \eqref{eq4.10} follows from Theorem~\ref{thm4.2} applied to the
critical energy $E_0 = n^2\pi^2$. If, on the other hand, $\omega_0 = 1$, then we conclude
in the same way, now based on the critical energy $E_0 = n^2\pi^2 +\lambda$.
\end{proof}

In the case where the $\omega_n$ are i.i.d.\ random variables, we can say even more
almost surely by invoking Theorem~\ref{dreiviertelschranke}. The condition $\eta_0 \not=
\eta_1 \mod \pi$ of Theorem~\ref{dreiviertelschranke} is fulfilled for all critical
energies (with $\eta_0, \eta_1$ now given by $k, \alpha$) throughout the
$\lambda$-interval $(0,\pi^2)$ under consideration.

\begin{coro} \label{randomstepcor}
If $\lambda$, $g_0$, $g_1$ and $H_{\omega}$ are as above, and the $\omega_n$ are i.i.d.\
random variables, then
\begin{equation} \label{randomstepbound}
\beta_f^-(p) \ge p-1/2
\end{equation}
for almost every $\omega$ and any $f\in L^2(0,1)$, $f\not= 0$.
\end{coro}

The previous corollary is particularly interesting as in this case it was proven in
\cite{DSS} that the operator $H_{\omega}$ given by \eqref{eq4.1a}, \eqref{eq4.1b} almost
surely exhibits pure point spectrum  with exponentially decaying
eigenfunctions, assuming only that $g_0 \not= g_1$. Thus the case $g_0=0$, $g_1= \lambda
\chi_{[0,1]}$ gives an example of a continuum random Schr\"odinger operator with
coexistence of spectral localization and super-diffusive transport ($\beta_f^-(2) \ge
3/2$).

Also, \cite{DSS} establishes dynamical localization for
$H_{\omega}$ in the following sense: If $g_0 \not= g_1$, then
there is a discrete set $M \subset \R$ such that for every compact
interval $I \subset \R \setminus M$, every compact set $K \subset
\R$, and every $p>0$,
\begin{equation} \label{eq4.13}
\E \left\{ \sup_{t\in\R} \| |X|^{p/2} e^{-itH_{\omega}} P_I(H_{\omega}) \chi_K \|
\right\} < \infty,
\end{equation}
where $P_I$ is the spectral projection onto $I$. Corollary~\ref{randomstepcor} shows that
the insertion of $P_I(H_{\omega})$ is crucial here: $\E \{ \sup_t \| |X|^{p/2}
e^{-itH_{\omega}} \chi_{[0,1]} \| \} < \infty$ would imply that $\beta_f^-(p) =0$ for
almost every $\omega$ and every $f$ supported in $[0,1]$, contradicting
\eqref{randomstepbound} if $p>1/2$. Thus, dynamical localization holds for the model
\eqref{eq4.1a}, \eqref{eq4.1b} in general only away from a discrete set of critical
energies.

\section{Self-Similar Potentials}

In this section, we discuss operators whose potentials are generated by means of a
substitution rule. The inherent self-similar structure of such potentials is expressed by
the existence of a renormalization scheme that gives rise to a dynamical system, the
so-called trace map, which governs the evolution of transfer matrix traces along the
different levels of the hierarchy. Results on the dynamics of the trace map can often be
used to establish power-law bounds for the norms of transfer matrices associated with
suitable energies. In the discrete case, three prominent models were studied in
\cite{DT}, namely, the Fibonacci model, the period doubling model, and the Thue-Morse
model. The strongest dynamical bound was obtained for the Thue-Morse model. We shall
carry out an explicit analysis for continuum operators with Thue-Morse and period
doubling symmetry, obtaining the same quantitative bounds, and then discuss the Fibonacci
case briefly.

The Thue-Morse substitution on the alphabet $\{ a, b \}$ is given by $S(a) = ab$, $S(b) =
ba$. This mapping extends to words over this alphabet by concatenation. Thus, for
example, $S^2(a) = abba$, $S^3(a) = abbabaab$. Let $\Omega_{{\rm TM}}$ be the associated
(two-sided) subshift, that is,
$$
\Omega_{{\rm TM}} = \{ \omega \in \{ a,b \}^\Z : \text{ every subword of $\omega$ is
contained in $S^n(a)$ for some } n \in \Z_+ \}.
$$
Now choose two numbers $l_a,l_b > 0$ and two local potentials $V_a \in L^1(0,l_a)$ and
$V_b \in L^1(0,l_b)$. Each sequence $\omega \in \Omega_{{\rm TM}}$ generates a potential
on $\R$ by
$$
V_\omega (x) = V_{\omega_0}(x) \text{ on } (0,l_{\omega_0}) , \; \; V(x) = V_{\omega_1}(x
- l_{\omega_0}) \text{ on } (l_{\omega_0}, l_{\omega_0} + l_{\omega_1}) , \text{ etc.},
$$
and similarly on the left half-line, using $\{ \omega_j \}_{- \infty < j \le -1}$.

\begin{theorem}\label{tmtheo}
For every pair $(V_a, V_b)$, there are $E_0 \in \R$ and $C > 0$ such that for every
$\omega \in \Omega_{{\rm TM}}$,
\begin{equation}\label{tmtmbound}
\| M_\omega (x,y,E_0) \| \le C \text{ for all } x,y \in \R.
\end{equation}
\end{theorem}

\begin{proof}
If $x_1 \ldots x_n$ is a word over the alphabet $\{ a,b \}$ and $E \in \R$, we denote by
$M(x_1 \ldots x_n,E)$ the transfer matrix over an interval of length $l_{x_1} + \cdots +
l_{x_n}$ with potential given by $V_{x_1} \cdots V_{x_n}$ and energy $E$. Define
$$
M^{(0)}_k (E) = M(S^k(0),E), \;\; M^{(1)}_k (E) = M(S^k(1),E),
$$
and
$$
x_k(E) = {\rm tr} \, M^{(0)}_k (E) , \; y_k(E) = {\rm tr} \, M^{(1)}_k(E).
$$
It is clear that $x_k = y_k$ for $k \ge 1$ and it follows from the substitution rule that
$$
M^{(0)}_{k}(E) = M^{(1)}_{k-1}(E) M^{(0)}_{k-1}(E) , \; M^{(1)}_{k}(E) = M^{(0)}_{k-1}(E)
M^{(1)}_{k-1}(E)
$$
and
\begin{equation}\label{tmtm}
x_{k}(E) = x_{k-2}(E)^2 (x_{k-1}(E) - 2) + 2 \; \mbox{ for } k \ge 3.
\end{equation}
The relation \eqref{tmtm} is called the Thue-Morse trace map.

Fix some $k \ge 3$ and let $\mathcal{E}_k = \{ E : x_{k-2}(E) = 0 \}$. It follows from
Floquet theory that $\mathcal{E}_k$ is countably infinite. We claim that for every $E \in
\mathcal{E}_k$,
\begin{equation}\label{ident}
M^{(0)}_k(E) = M^{(1)}_k(E) = \left( \begin{array}{rr} 1 & 0 \\ 0 & 1 \end{array}
\right).
\end{equation}
This is a consequence of the Cayley-Hamilton theorem:
\begin{eqnarray*}
M^{(0)}_{k}(E) & = & M^{(0)}_{k-2}(E) \, M^{(1)}_{k-2}(E) \, M^{(1)}_{k-2}(E) \, M^{(0)}_{k-2}(E) \\
& = & M^{(0)}_{k-2}(E) \left( x_{k-2}(E) M^{(1)}_{k-2}(E) - I \right) M^{(0)}_{k-2}(E) \\
& = & - M^{(0)}_{k-2}(E) M^{(0)}_{k-2}(E) \\
& = & - \left( x_{k-2}(E) M^{(0)}_{k-2}(E) - I \right) \\
& = & I
\end{eqnarray*}
and, similarly, $M^{(1)}_k(E) = I$. This yields \eqref{ident}. From this,
\eqref{tmtmbound} follows readily.
\end{proof}

Thus, we can apply Corollary~\ref{oneenergy} with $\alpha = 0$ and obtain
$\beta_{\omega,f}^- (p) \ge p - 1$ for suitable compactly supported $f$. Note that we can
consider operators either on the half-line or on the whole line; the respective version
of Corollary~\ref{oneenergy} then tells us what is required from $f$.

Next we consider the period doubling substitution on the alphabet $\{ a, b \}$, which is
given by $S(a) = ab$, $S(b) = aa$. Again, we define the associated (two-sided) subshift
$\Omega_{{\rm PD}}$ and choose two local potentials $V_a \in L^1(0,l_a)$ and $V_b \in
L^1(0,l_b)$, generating potentials $V_\omega$ as before.

\begin{theorem}\label{pdtheo}
For every pair $(V_a, V_b)$, there are $E_0 \in \R$ and $C > 0$ such that for every
$\omega \in \Omega_{{\rm PD}}$,
\begin{equation}\label{pdtmbound}
\| M_\omega (x,y,E_0) \| \le C (1 + |x - y|) \text{ for all } x,y \in \R.
\end{equation}
\end{theorem}

\begin{proof}
Define
$$
M^{(0)}_k (E) = M(S^k(0),E), \;\; M^{(1)}_k (E) = M(S^k(1),E),
$$
and
$$
x_k(E) = {\rm tr} \, M^{(0)}_k (E) , \; y_k(E) = {\rm tr} \, M^{(1)}_k(E).
$$
It follows from the substitution rule that
$$
M^{(0)}_{k}(E) = M^{(1)}_{k-1}(E) M^{(0)}_{k-1}(E) , \; M^{(1)}_{k}(E) = M^{(0)}_{k-1}(E)
M^{(0)}_{k-1}(E)
$$
and
\begin{equation}\label{pdtm}
x_{k}(E) = x_{k-1}(E) y_{k-1}(E) - 2, \;\; y_{k}(E) =  (x_{k-1}(E))^2 - 2.
\end{equation}
The relation \eqref{pdtm} is called the period doubling trace map.

Fix some $k \ge 2$ and let $\mathcal{E}_k = \{ E : x_{k-1}(E) = 0 \}$. It follows again
from Floquet theory that $\mathcal{E}_k$ is countably infinite. For $E \in
\mathcal{E}_k$, \eqref{pdtm} yields $x_k(E) = - 2$. Thus, there is a constant $\gamma$
such that
$$
M^{(0)}_k(E) \text{ is conjugate to } \left( \begin{array}{rr} -1 & \gamma \\
0 & -1 \end{array} \right).
$$
Moreover, it follows from the Cayley-Hamilton theorem that
$$
M^{(1)}_k(E) = M^{(0)}_{k-1}(E) M^{(0)}_{k-1}(E) = x_{k-1}(E) M^{(0)}_{k-1}(E) -
\text{Id} = - \text{Id} .
$$
The bound \eqref{pdtmbound} is now an immediate consequence of these two observations.
\end{proof}

Thus, we can apply Corollary~\ref{oneenergy} with $\alpha = 1$ and obtain
$\beta_{\omega,f}^- (p) \ge (p - 5)/2$ for suitable compactly supported $f$.

We conclude this section with a brief discussion of the Fibonacci case. The Fibonacci
substitution on the alphabet $\{ a, b \}$ is given by $S(a) = ab$, $S(b) = a$. As before,
we may define the subshift $\Omega_F$ generated by the substitution and, given two local
potentials, a family of Schr\"odinger operators. It is possible to prove power-law upper
bounds for the associated transfer matrices for suitable energies. This is technically
much more involved than in the Thue-Morse or period doubling case, but it can be
accomplished using ideas from \cite{DT,DT2,IT,KKL}. The analysis in those papers is to a
large extent independent of the explicit form of the transfer matrices and is mainly
based on the renormalization scheme that arises naturally from the substitution rule.

The Fibonacci case is different from Thue-Morse and period doubling on a conceptual level
as there are no ``exceptional'' enegies. In fact, one can prove power-law bounds for all
energies in the spectrum. In the discrete case, one can even choose the power uniformly
on the spectrum. Thus, the methods in \cite{DT} and this paper should not be expected to
give the best dynamical results in the Fibonacci case. Indeed, the best known dynamical
results for the discrete version of the Fibonacci potential are contained in \cite{DT2}.
The latter paper combines ideas from \cite{DT} and \cite{JL1,KKL} and gives quite strong
dynamical bounds in cases where one has quite good solution estimates.

\section{Further remarks} \label{Remarks}

In this section we address a number of issues that are suggested by our work. Most
importantly, we extend the main result in the discrete case from \cite{DT} to more
general finitely supported initial states.

\subsection{Dynamical Bounds for Discrete Schr\"odinger Operators}

Given a bounded $V : \Z \to \R$, we may consider the discrete Schr\"odinger operator
$$
[H\varphi](n) = \varphi(n+1) + \varphi(n-1) + V(n) \varphi(n)
$$
on $\mathcal{H} = \ell^2(\Z)$ or $\ell^2(\Z_+)$, $\Z_+ = \{1,2,\ldots\}$, (with a
suitable boundary condition at the origin; e.g., Dirichlet) and the associated difference
equation
\begin{equation}\label{eved}
u(n+1) + u(n-1) + V(n) u(n) = E_0 u(n).
\end{equation}
The particular solution of \eqref{eved} satisfying $u(0) = 0$, $u(1) = 1$ will be denoted
by $u_{0,E_0}$.

The position operator acts as $[X\varphi](n) = n \varphi(n)$ and we can define the
quantities $M_f(T,p)$, $M_{f,D} (T,p)$, $\beta_f^-(p)$, and $\beta_{f,D}^-(p)$ as before;
compare \eqref{eq1.3} and \eqref{eq1.4}.

Transfer matrices and the sets $P(\alpha,C,N)$ are also defined in a completely analogous
way. With the standard scalar product on $\mathcal{H}$, we may now state the following
pair of results, which are the discrete analogs of Theorems~\ref{mainhl} and
\ref{mainwl}.

\begin{theorem}\label{mainhld}
Suppose $E_0\in \R$ is such that there exist $C>0$ and $\alpha >0$ with $E_0\in
P(\alpha,C,N)$ for all sufficiently large $N$. Let $A(N)$ be a subset of $P(\alpha,C,N)$
containing $E_0$ such that $\mathrm{diam} (A(N))\longrightarrow 0$ as $N \to \infty$.
Then, for every finitely supported $f \in \ell^2 (\Z_+)$ with $\langle u_{0,E_0},
f\rangle \neq 0$, there exists $\widetilde{C}>0$ such that for $T$ large enough, $M_{f,D}
(T,p)  \geq \widetilde{C}  |B(T)| T^{\frac{p - 3\alpha}{1 + \alpha}}$, where $B(T)$ is
the $1/T$ neighborhood of $A(T^{\frac{1}{1+\alpha}})$.
\end{theorem}

\begin{theorem}\label{mainwld}
Suppose $E_0\in \R$ is such that there exist $C>0$ and $\alpha >0$ with $E_0\in
P(\alpha,C,N)$ for sufficiently large $N$. Let $A(N)$ be a subset of $P(\alpha,C,N)$
containing $E_0$ such that $\mathrm{diam} (A(N))\longrightarrow 0$ as $N \to \infty$. Let
$f \in \ell^2 (\Z)$ be finitely supported and satisfy $\langle u, f\rangle \neq 0$ for at
least one solution $u$ of \eqref{eved}. Then, for $T$ large enough, $M_f (T,p)  \geq
\widetilde{C}  |B(T)| T^{\frac{p - 3\alpha}{1 + \alpha}}$, where $B(T)$ is the $1/T$
neighborhood of $A(T^{\frac{1}{1+\alpha}})$.
\end{theorem}

These results are proved in the exact same way as their continuum counterparts. When
specializing Theorems~\ref{mainhld} and \ref{mainwld} to the case $f = \delta_1$, we
recover the results of Damanik and Tcheremchantsev from \cite{DT} (for isolated critical
energies). Notice that the assumption $\langle u_{0,E_0}, f\rangle \neq 0$ (resp.,
$\langle u, f\rangle \neq 0$ for at least one solution $u$ of \eqref{eved}) is trivially
satisfied in this case and hence was not an issue in \cite{DT}. This is also the reason
why the results of \cite{DT} do not immediately suggest the correct formulation of an
extension to more general initial states.

Let us discuss the example of a (random) dimer model in more detail. On the one hand,
this will generalize results of \cite{JSBS} and, on the other hand, this will provide the
discrete analog of our discussion in Section~\ref{sec5}; particularly,
Corollaries~\ref{cor4.3} and \ref{randomstepcor}. A dimer model is a discrete
Schr\"odinger operator on the whole line whose potential $V$ takes values in the set $\{
\lambda, -\lambda \}$, $\lambda > 0$, and satisfies $V(2n) = V(2n-1)$ for all $n$. A
random dimer model is a family of dimer models $\{H_\omega\}$, where $\omega \in \{
\lambda, -\lambda \}^\Z$, $V_\omega(2n) = \omega_n$, and the $\omega_n$'s are i.i.d.\
random variables. Notice that the energies $E_0 = \pm \lambda$ are critical if $0 <
\lambda < 1$. It is straightforward to see that Theorem~\ref{mainwld} above, combined
with \cite[Theorem~7]{JSBS}, implies the following:

\begin{coro}
Let $\lambda \in (0,1)$ and $f \not= 0$ be supported in $\{1,2\}$. Then, for every $p
> 0$,
\begin{itemize}
\item[{\rm (i)}] $\beta_{f,\omega}^-(p) \ge p-1$ for every
$\omega$, \item[{\rm (ii)}] $\beta_{f,\omega}^-(p) \ge p-1/2$ for
almost every $\omega$.
\end{itemize}
\end{coro}

A straightforward calculation shows that the condition $\eta_0-\eta_1 \not= 0 \mod \pi$,
required in \cite[Theorem~7]{JSBS}, holds at $E_0=\pm\lambda$ for every $\lambda \in
(0,1)$.

For $f = \delta_1$, the above was shown in \cite{JSBS}. Here we
see that, by translation invariance, we may in fact take all
non-trivial initial states $f$ that have their support in one of
the random blocks. In other words, this is the precise analog of
Corollaries~\ref{cor4.3} and \ref{randomstepcor} from
Section~\ref{sec5}.

\subsection{Open Problems}

We conclude this paper with a discussion of open problems that are suggested by our work
and previous ones.

Our dynamical results require a certain non-orthogonality
condition from the initial state $f$. Such a condition can
certainly not be dropped in general as the Bernoulli-Anderson
example shows: If $f$ has no energy near a critical one, the
results of \cite{DSS} show that no non-trivial dynamical lower
bound exists. This suggests studying cases where $\langle
u_{0,E_0}, f\rangle = 0$, $E_0$ critical, but $\langle u_{0,E},
f\rangle \neq 0$ for energies $E$ close to $E_0$. For example, is
it possible to prove a non-trivial dynamical lower bound if the
function $E \mapsto \langle u_{0,E}, f\rangle$ vanishes to a
finite order at $E_0$? (Note that all roots of $\langle
u_{0,E},f\rangle$ are of finite order if $f$ is compactly
supported and non-zero: In this case $\{u_{0,E}:E\in\R\}$ is total
in $L^2$ over the support of $f$ and thus the analytic function
$\langle u_{0,E}, f\rangle$ doesn't vanish identically.)

Our results for the Bernoulli-Anderson model once again motivate the following question:
Is the bound \eqref{eq1.7} optimal? For the random dimer model, it was conjectured in
\cite{DWP} that indeed $\beta_{\delta_1,\omega}^-(2) = 3/2$ for almost every $\omega$.
That $3/2$ is a lower bound was shown in \cite{JSBS}, and here we proved an analogous
result for the Bernoulli-Anderson model with a critical energy. Proving dynamical upper
bounds, especially on moments of the position operator, is a hard problem. The only
existing results in this direction for random Schr\"odinger operators\footnote{For a
class of sparse potentials, Tcheremchantsev has explicitly determined the diffusion
exponents \cite{T}.} establish complete dynamical localization in the sense that all
diffusion exponents vanish. It is not clear how to deal with critical energies in terms
of proving dynamical upper bounds, and we consider this an interesting open problem.

\appendix

\section{Existence of analytic eigenvectors} \label{appendixA}

The following fact from Floquet theory has been used in the proof
of Lemma~\ref{periodic} above. This is probably well known. We
include a proof mainly for the reason that we use it not only for
energies $E_c$ in the interior of stability intervals (where
$|D_0(E_c)|<2$), but also at degenerate band edges ($E_c$ in the
interior of the spectrum, but $|D_0(E_c)|=2$). For general
background on Floquet theory see \cite{Eastham}.

\begin{lemma} \label{analyticev}
Let $H^{(0)}$ be a periodic Schr\"odinger operator as in
Section~\ref{deterministic} and $E_c$ in the interior of
$\sigma(H^{(0)})$. Then there exists an open neighborhood $I$ of
$E_c$ and analytic functions $\rho_{\pm}:I\to \C$ and $v_{\pm}:I
\to \C^2$ such that for each $E\in I$, $T_0(E) v_{\pm}(E)=
\rho_{\pm}(E)v_{\pm}(E)$, the $v_{\pm}(E)$ are linearly
independent, and $\rho_-(E) = \overline{\rho_+(E)}$,
$v_-(E)=\overline{v_+(E)}$.
\end{lemma}

\begin{proof}
Define $D_0 (E) :=\tr\, T_0 (E)$. As $E_c$ belongs to the spectrum
of $H^{(0)}$, we have
$$
-2 \leq D_0 (E)  \leq 2.
$$

We consider two cases:

\smallskip

\textit{Case 1:} $-2 < D_0 (E_c) < 2$.

\smallskip

Then, $-2 < D_0 (E) < 2$ for all $E$ in an interval $I$ around
$E_c$.  In this interval $T_0 (E)$ has the different complex
conjugate eigenvalues
\begin{equation} \label{appeval}
\rho_{\pm} (E) = \frac{1}{2} (D_0 (E) \pm i \sqrt{ 4 - D_0 (E)^2})
\end{equation}
with corresponding linearly independent complex conjugate
eigenvectors
\begin{equation} \label{appevec}
v_{\pm}(E) = \left( \begin{array}{c}  1\\
c_{\pm} (E)
\end{array} \right).
\end{equation}
Here, $\sqrt{} \, :\, [0,\infty) \longrightarrow [0,\infty)$ is
the usual square root and
\begin{equation} \label{appc}
c_{\pm} (E) = \frac{ \rho_{\pm} (E) - u_N (1,E)}{u_D (1,E)}
\end{equation}
where $u_N$ and $u_D$ are the solutions of $-u'' + V_0 u = E u $
with initial conditions $u_N (0) = u_D ' (0) =1$ , $u_N' (0) = u_D
(0) = 0$. This is well known and easily checked. In particular,
$u_D(1,E)\not= 0$ as $E$ is not an eigenvalue of the Dirichlet
problem on $[0,1]$.

\medskip

\textit{Case 2:}  $|D_0 (E_c)|=2$.

As $E_c$ is in the interior of the spectrum of $H^{(0)}$, we are
at a degenerate band edge. Thus, $T_0 (E_c)$ is equal to
$\mbox{id}$ or $-\mbox{id}$.

Assume w.l.o.g.\ $D_0 (E_c) = 2$.  As $E_c$ belongs to the interior of the spectrum of
$H^{(0)}$, $D_0$ has a local maximum at $E_c$, $D_0'(E_c) =0$. As local extreme values of
$D_0$ are non-degenerate (see, e.g., Section~2.3 of \cite{Eastham}, in particular
page~29), $D_0''(E_c)<0$. This implies that $\sqrt{2-D_0(E)}$ and therefore
$\sqrt{4-D_0(E)^2}$ have a branch which is analytic in a neighborhood $I$ of $E_c$. We
now use this branch in the definition of $\rho_{\pm}(E)$ via \eqref{appeval}. Thus
$\rho_{\pm}(E)$ are analytic and complex conjugate eigenvalues of $T_0(E)$ near $E_c$.
One checks that
\begin{equation} \label{apprhoderiv}
\rho_{\pm}'(E_c) = \pm i\sqrt{c}\not= 0,
\end{equation}
where $c=|D_0''(E_c)|/2$. We again define $c_{\pm}$ and $v_{\pm}$ by \eqref{appc} and
\eqref{appevec}, which makes them analytic up to a possible singularity at $E_c$.
However, both $\rho_{\pm} (E) - u_N (1,E)$ and $u_D (1,E)$ have first order zeros at
$E_c$. For the former, this follows from \eqref{apprhoderiv}, noting that $u_N(1,E)$ is
real. For the latter, this can be seen by using that the Pr\"ufer phase
\begin{equation} \label{apppruefer}
\theta(1,E) := \arctan \frac{u_D(1,E)}{u_D'(1,E)}
\end{equation}
has positive $E$-derivative (one way to prove this may be found in
\cite[Section~4]{Stolz}). Using that $u_D(1,E_c)=0$ and $u_D'(1,E_c)=1$ it is now easily
verified from \eqref{apppruefer} that $(\partial_E u_D)(1,E_c) = (\partial_E
\theta)(1,E_c) >0$.

We conclude that the singularity of $c_{\pm}$ at $E=E_c$ is
movable. We also see, from l'Hospital's rule, that $c_{\pm}(E_c)$
has non-vanishing imaginary part. We conclude that for all $E$
near $E_c$ the $v_{\pm}(E)$ are eigenvectors of $T_0(E)$ (this is
trivial for $E=E_c$), which are complex conjugate and linearly
independent.
\end{proof}

\end{document}